\documentclass[10pt,conference]{IEEEtran} 
\usepackage{cite}
\usepackage{amsmath,amssymb,amsfonts}
\usepackage{algorithmic}
\usepackage{graphicx}
\usepackage{textcomp}
\def\BibTeX{{\rm B\kern-.05em{\sc i\kern-.025em b}\kern-.08em
    T\kern-.1667em\lower.7ex\hbox{E}\kern-.125emX}}

\newcommand{\fit}{\textsc{fit}}

\begin{document}
\title{Extracting Fractional Inspiratory Time from Electrocardiograms}

\author{\IEEEauthorblockN{Maria Nyamukuru}
\IEEEauthorblockA{\textit{Thayer School of Engineering} \\
\textit{Dartmouth College}\\
Hanover, NH, USA \\
maria.t.nyamukuru.th@dartmouth.edu}
\and
\IEEEauthorblockN{Kofi Odame}
\IEEEauthorblockA{\textit{Thayer School of Engineering} \\
\textit{Dartmouth College}\\
Hanover, NH, USA \\
kofi.m.odame@dartmouth.edu}
}

\maketitle

\begin{abstract}
Non-invasive at-home monitoring of lung and lung airways health enables the early detection and tracking of respiratory diseases like asthma and chronic obstructive pulmonary disease (COPD). Various proposed approaches estimate the respiratory rate and extract the respiratory waveform from an electrocardiogram (ECG) signal as a way to discreetly monitor lung health. Unfortunately, these approaches fail to accurately capture the respiratory cycle phase features, resulting in a non-specific, incomplete picture of lung health. This paper introduces an algorithm to extract more respiratory information from the ECG signal by framing the problem as a binary segmentation task. In addition to respiratory rate (RR), the algorithm derives the fractional inspiratory time (\fit), a direct measure of airway obstruction derived from respiratory phase information. The algorithm is based on a gated recurrent neural network that infers vital respiratory information from a single-lead ECG signal. We measure our algorithm's performance on 5 subjects from the MIMIC dataset and 5 subjects from the CEBS database. Our algorithm maintains exceptional performance in estimating the respiratory rate and outperforms current algorithms that extract the respiratory cycle phases and \fit/ inspiratory:expiratory ratio (IER). Our algorithm reports a root mean squared error (RMSE) of 0.06 in the computation of \fit~(values range from 0.2-0.6) and a RMSE of 0.54 breaths per minute (bpm) for respiratory rate (values range from 8 - 28 breaths per minute (bpm)) on the MIMIC dataset, and an \fit~RMSE of 0.11 and and RR RMSE of 0.66 bpm on the CEBS dataset.
\end{abstract}


\section{Introduction}
\label{sec:introduction}
There is a growing infrastructure of smartwatches, smartphones and wearable devices that can measure heart activity via electrocardiography (ECG) or photoplethysmography (PPG) \cite{gaurav2016cuff, koltowski2019kardia, apple_watch_ecg_2020, kinnunen2020feasible}. Interestingly, this network of personal cardiac sensors is also an excellent platform for remote monitoring of lung health, because respiration affects the rhythm and electrical activity of the heart. As such, a lot of recent work has proposed ways to extract breathing rate from ECG and PPG signals \cite{7997854, 6497470, Lzaro2014ElectrocardiogramDR, 5966727, 7873222, 4811954, Sharma2018ECGderivedRB}.

\begin{figure}[ht!]
\centerline{\includegraphics[scale=0.12]{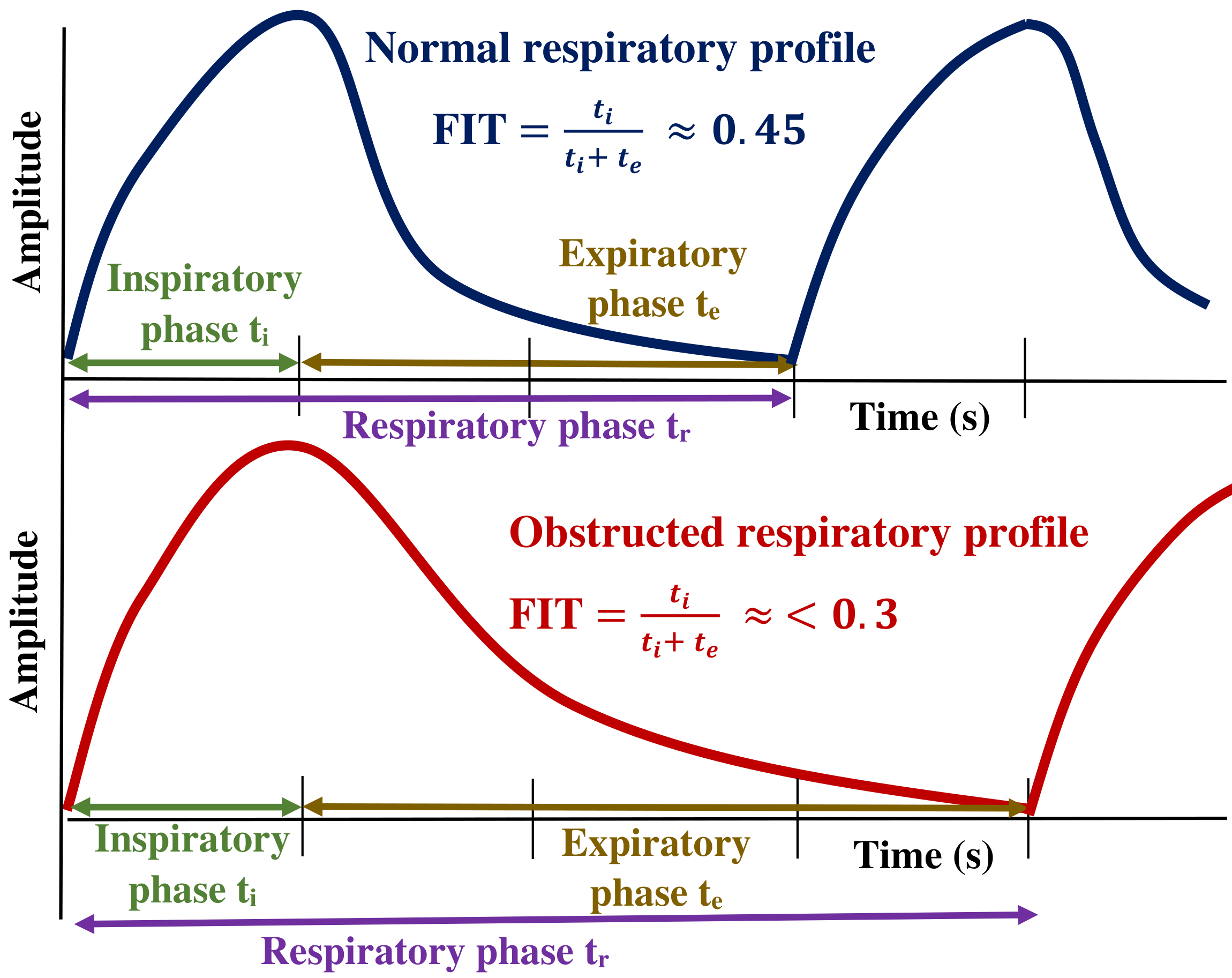}}
\caption{Top panel: normal respiratory signal. Bottom panel: respiratory signal in lung airway obstruction, typified by an elongated expiratory phase.}
\label{obstructed_signal}
\end{figure}

Unfortunately, while breathing rate can provide a coarse picture of lung health, it is inadequate for inferring more detailed information, like the extent and severity of  lung airway obstruction. Monitoring lung airway obstruction---a condition where inflammation and excess mucus production in the lungs' airways results in impeded airflow---is critical for the early detection of symptom flare-ups in chronic obstructive pulmonary disease (COPD) \cite{Watz2018SpirometricCD}, \cite{Berge2012PredictionAC}. If we detected and treated these COPD flare-ups earlier, it would increase patients' quality of life \cite{Wilkinson2004EarlyTI}, reduce mortality \cite{Ho2014InHospitalAO, Suissa2012LongtermNH}, and eliminate billions of dollars in COPD healthcare costs \cite{Guarascio2013TheCA}.

\begin{figure*}
\centerline{\includegraphics[scale=0.55]{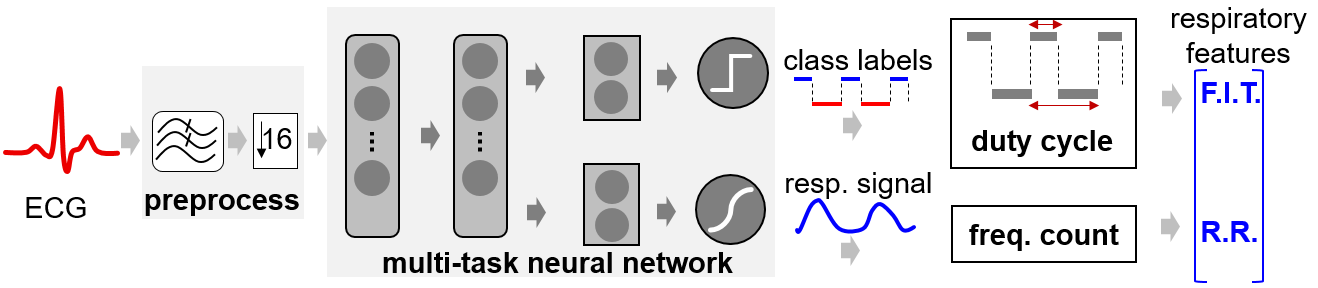}}
\caption{Block diagram of the proposed algorithm. A GRU network classifies the ECG signal as belonging to either the inspiratory or expiratory phases of breathing, and \fit~is derived from the resulting class labels, via a duty cycle calculation. The GRU network also outputs an estimate of the respiratory signal, from which the respiratory rate is extracted.}
\label{blockdiagram}
\end{figure*}

A few researchers have gone beyond breathing rate extraction to introduce algorithms that can analyze ECG/PPG signals for more direct measures of lung airway obstruction \cite{Prinable2020DerivationOR, 7408587}. This work intends to expand on the existing body of work to extract more breathing metrics from ECG signals. In this paper, we present an algorithm to accurately extract much finer details about the respiratory pattern from ECG with higher accuracy than previously achieved in \cite{Prinable2020DerivationOR, 7408587}. Specifically, our algorithm extracts the duration of the inspiratory and expiratory phases of the respiratory cycle, by analyzing the temporal dependencies in ECG over multiple time scales with a gated recurrent neural network. Deriving such detailed respiratory features from the cardiac signal would allow us to classify the severity of airway obstruction with over $96\%$ accuracy \cite{odame2020inferring}.

\section{Effect of Airway Obstruction on ECG}
A person experiencing lung airway obstruction will take longer than usual to exhale \cite{Tobin1983BreathingP2, doi:10.1164/ajrccm-conference.2017.195.1_MeetingAbstracts.A3007, Miki2018ExerciseTA}. This is reflected in an abnormally short \emph{fractional inspiratory time} (\fit), which is the duration of the inspiratory phase as a fraction of the total respiratory period (see Fig.  \ref{obstructed_signal}) \cite{kaplan2000detection}. \fit~drops from a normal range of $0.45-0.5$ to as low as $0.2$ in the presence of severe airway obstruction \cite{Tobin1983BreathingP2, doi:10.1164/ajrccm-conference.2017.195.1_MeetingAbstracts.A3007, Miki2018ExerciseTA}.
Such alterations to the breathing pattern in turn produce subtle changes to the ECG signal. For instance, the amplitude envelope of the ECG's R peaks is modulated during the breathing cycle, due to changes in the heart's position as the diaphragm expands and contracts \cite{5333113}. When airway obstruction increases the exhalation time, this is reflected by a longer positive half cycle in the ECG's amplitude modulation envelope.
Also, the heart rate increases during inhalation and decreases during exhalation due to pressure changes in the thoracic cavity \cite{826020}. So, the longer-than-usual exhalation times caused by airway obstruction will produce corresponding longer periods of lowered heart rate that are observable in the ECG signal.

\section{Related Work}
Many algorithms exist for extracting the respiratory rate from either ECG or pulse oximetry data  \cite{7873222, 4811954, Sharma2018ECGderivedRB}. But respiratory rate alone does not provide clinically-relevant information about the patient's pulmonary health. Much more informative than respiratory rate are the relative durations of the inspiratory and expiratory (I/E) phases of breathing; lung airway obstruction causes the expiratory phase to lengthen relative to the inspiratory phase (Fig. \ref{obstructed_signal}). 

In an attempt to extract useful I/E information from pulse oximetry data, Prinable et al. \cite{Prinable2020DerivationOR} explored LSTM and U-Net algorithms to extract breathing signal and respiratory metrics like respiratory rate and IER from the PPG signals of 11 healthy subjects and 11 subjects with asthma breathing at a range of controlled respiratory rates. The neural networks produced an estimate of the respiratory signal, from which the I/E phase durations were calculated as the distances between peaks and troughs. Unfortunately, the regression task poorly captures the asymmetry of a true respiratory signal and thus reports low pearson correlation coefficients ($R^2$) for inspiratory time (0.69), expiratory time (0.47), and IER (0.01). 

A different approach for extracting I/E information from ECG was proposed by Alamdari et al. \cite{7408587}. In that algorithm, the respiratory signal was estimated as the envelope of the ECG, and the respiratory cycle phases from 19 healthy subjects were calculated as occurring between the envelope peaks and troughs. Similar to Prinable et al. \cite{Prinable2020DerivationOR}, the ECG envelope mainly captured the respiratory signal's fundamental frequency. As such, the Alamdari algorithm's ability to distinguish inhalation from exhalation using ECG signals is only 67.2 \% accurate \cite{7408587}. Such a low accuracy would produce large errors in calculating the relative durations of the inspiratory and expiratory phases.

Unlike previous work, our approach for extracting I/E information does not depend critically on the peak and trough locations of a cardiac-derived respiratory signal. Instead, we have framed the problem as a binary classification task: our algorithm learns to infer whether a given ECG data sample occurred during inhalation or exhalation.

\section{Proposed Algorithm}

Fig. \ref{blockdiagram} shows a block diagram our algorithm for inferring I/E respiratory phase information from ECG data. The only input to the algorithm is raw ECG data and no other patient-related features. The outputs are the respiratory rate and the fractional inspiratory time. In this section, we discuss the main components of the algorithm, which comprises a neural network, a duty cycle calculator and a frequency detector. 

\subsection{Multi-task Neural Network} 
\label{MTL_sec}
The first stage of our algorithm is a neural network to identify whether an input ECG data sample occurred during breath inhalation or exhalation. To capture the multi-scale temporal dependencies that are inherent to this task---ECG is a high frequency signal compared to the respiratory signal---we use a gated recurrent unit (GRU) neural network. In addition to identifying long-term temporal dependencies in the ECG signal, a GRU is also resource-efficient; it requires no data buffer, needing only to process one data sample at a time, while embedding past information within its hidden states \cite{sherstinsky2020fundamentals}. This makes it suitable to use in real-time settings on streaming data for immediate feedback.

\begin{figure}[t]
\hspace*{-0.5cm}
\centerline{\includegraphics[scale=0.35]{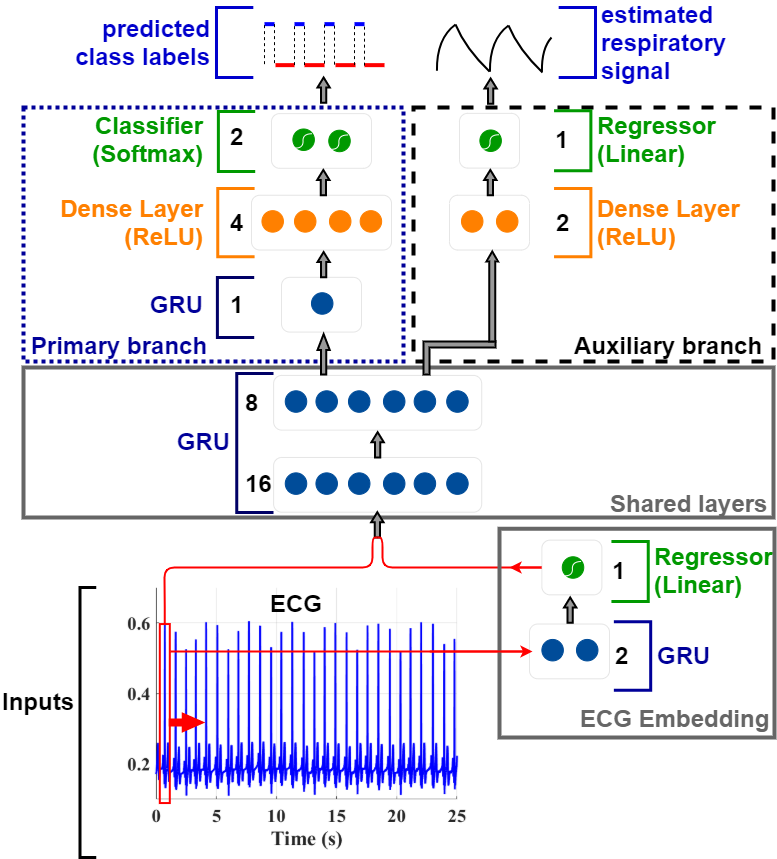}}
\caption{Two-parallel-branch multi-task learning neural network model that takes in 2 features at a time (ECG and embedded ECG signal), and outputs both the respiratory phase classes (primary branch), and respiratory signal (auxiliary branch) for each sample in a 25-second window with no overlap.}
\label{gru_nn}
\end{figure}

Fig. \ref{gru_nn} shows our neural network architecture, which is a two-parallel-branch multi-task learning (MTL) GRU model. The MTL model comprises a few initial shared layers, followed by two separate branches that each perform a specific function. One branch of the GRU network handles the primary task, which is the binary classification (inhalation or exhalation) of a given ECG sample. The \fit~is computed from the output of this branch. The second branch deals with a secondary/ auxiliary task: to estimate the respiratory signal from the ECG signal. The respiratory rate (RR) is estimated from the output of this branch. The two tasks (primary and auxiliary) are related, since the inhalation/exhalation classes predicted in the primary task, are simply the phases of the respiratory signal, which is also being predicted in the auxiliary task. This suggests that there is a set of common ECG features that should work effectively across both tasks \cite{Sener2018MultiTaskLA, Argyriou2006MultiTaskFL, Baxter2000AMO}. The role of the shared layers is to therefore extract this set of relevant ECG features. 

Since the MTL model solves the binary classification problem while simultaneously estimating the respiratory signal, it is implicitly forced to make class predictions that are related to actual, underlying respiratory information (versus making predictions based on spurious artifacts). This approach produces a 16$\times$ increase in training speed (see Fig. \ref{jlearnvsnone}), compared to the single-task algorithm that is made of only the primary task.

\subsection{Duty Cycle}
\label{dcycle}
The duty cycle algorithm uses the following technique to estimate the \fit~in every window. First, the algorithm counts the total number of samples, $S_i$, that were classified as occurring during the inspiratory phase. Then, the estimated average fractional inspiratory time is simply $\fit=S_i/S_T$, where $S_T$ is the total number of samples in the window; $S_T=F_s \times w$ for $w$-second long window and a fixed sampling rate of $F_s$ Hz.


\subsection{Frequency Count}
\label{rrate}
The frequency count algorithm estimates the respiratory rate as RR = 60/$T_R$, where $T_R$ is the average period between peaks in the estimated respiratory signal.


\section{Experiments}
\subsection{Dataset}

We performed preliminary evaluation of our algorithm on data from the MIMIC database, which is publicly available on PhysioNet \cite{doi:10.1161/01.CIR.101.23.e215}. The database includes respiratory, ECG and PPG signals that were simultaneously collected in the intensive care unit (ICU) from patients with various illnesses, like respiratory failure and sepsis. To allow for timely dissemination of our results, we used the smaller MIMIC-I database \cite{moody1996database}, rather than the latest version (MIMIC-IV), which is almost 1000$\times$ larger and would require significantly more time and computing resources to process. Future work will evaluate our algorithm on the latest MIMIC database.

Our current study uses the ECG modified chest lead one (MCL1) signal as the algorithm's input, with the corresponding respiratory signal used as the target. The ECG MCL1 is similar to the ECG Lead 1 signal that is typically measured by a smartwatch; this allows us to explore the feasibility of implementing the \fit~extraction algorithm with a smartwatch or mobile device. There are 18 subjects in the MIMIC database with ECG MCL1 signals, each with one to several hours of recorded data. We separated these subjects into one group of 13 subjects for training and validation, and held out the remaining 5 subjects for final testing. More details on how this data was collected can be found here \cite{moody1996database}.

We also performed evaluation of the algorithm on data from the CEBS (Combined measurement of ECG, Breathing, and Seismocardiograms) database \cite{6713413}, publicly available on PhysioNet. This database comprises of ECG Lead I, Lead II and respiratory signals. For our study, we used the ECG Lead I signal as the algorithm's input, because it is the signal typically measured by a smartwatch. The signals in the CEBS database were collected using the Biopac MP36 data acquisition system, which ensured both simultaneous recording and temporal alignment of the ECG Lead I and respiratory signals. 

The CEBS database consists of one hour recordings of data from 20 subjects, all presumed healthy. We partitioned the subjects into two, with one set of 15 subjects used for training and validation, and the other 5 subjects used only for testing. More details on how this data was acquired can be found in \cite{6713413}. 

\subsection{Preprocessing}
The ECG and respiratory signals used in the training and validation datasets were partitioned into 25-second non-overlapping windows. This window size is enough to capture at least 3 respiratory periods, but not so long as to hamper training of the MTL neural network model. This resulted in 30304 and 1172 25-second windows in the MIMIC and CEBS respectively.

The ECG and respiratory signals used in the test dataset were partitioned into 60-second (1 minute) windows. This is to maintain the standard measurement units of breaths per minute (bpm) for the respiratory rate. This resulted in 982 and 300 60-second windows in the MIMIC and CEBS respectively.

After windowing, the signals were normalized to span the (0, 1) range. 

\subsection{Data Validation}
Since the MIMIC data was collected from the ICU, some 25-second windows of the respiratory signal are corrupted by motion artifact, making it impossible to establish ground truth. Data without ground truth labels cannot be used to train or even evaluate our neural network. We addressed this problem with data quality validation, discarding windows of data that display any of the following characteristics:
\begin{itemize}
    \item No peaks detected in the respiratory signal.
    \item Uncertain ground truth respiratory rate. We can calculate the ground truth respiratory rate with one of two different prominent peak detection algorithms. The first algorithm applies a peak detection algorithm to the respiratory signal, and measures the average period between successive peaks in order to compute the respiratory rate $\rm{RR}_1$. The second algorithm applies a peak detection algorithm to the auto-correlation sequence of the respiratory signal, and measures the average period between successive peaks in order to compute the respiratory rate $\rm{RR}_2$. If these two respiratory rate calculations differ by more than 100$\%$ for a given window, then the data is deemed too noisy. 
    \item Respiratory rate greater than 60 beats per minute.
    \item Spectral purity less than 0.9 dB. The respiratory signal is not a pure sinusoid, but it is also not completely aperiodic, lying somewhere in between these two extremes. Preliminary studies revealed that the ratio of the respiratory signal's fundamental frequency component to that of all other portions of the signal must be at least 0.9 dB for the peaks and valleys to be identified unambiguously. 
\end{itemize}

\begin{figure}[htb]
\centerline{\includegraphics[scale=0.63]{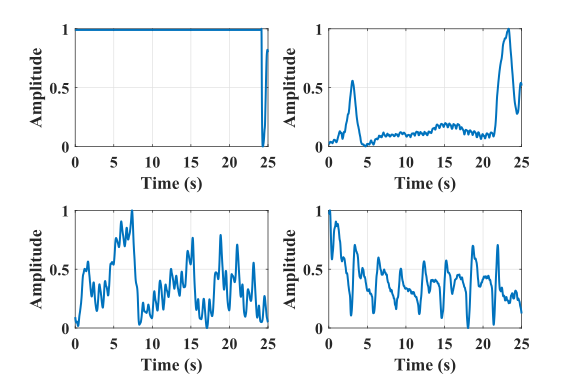}}
\caption{Four examples of unacceptable subject data in the MIMIC database, where no peaks are detected in the respiratory signal (top left), there is uncertainty in the ground truth respiratory rate (top right), respiratory rate is unknown and possibly greater than 60 bpm (bottom left), and the spectral purity is less than 0.9 dB (bottom right). The ground truth is impossible to establish, because there is uncertainty in the true locations of the peaks and valleys. Also, the atypical wave shapes suggest that these measurement have been severely corrupted by motion artifact. Without ground truth labels, it is impossible to use this data to train or even evaluate our neural network.}
\label{bad_subject}
\end{figure}

 Fig. \ref{bad_subject} shows some examples of respiratory signals that we flagged during the data validation process as being too noisy to ascertain ground truth.

\begin{figure}[htb]
\centerline{\includegraphics[scale=0.48]{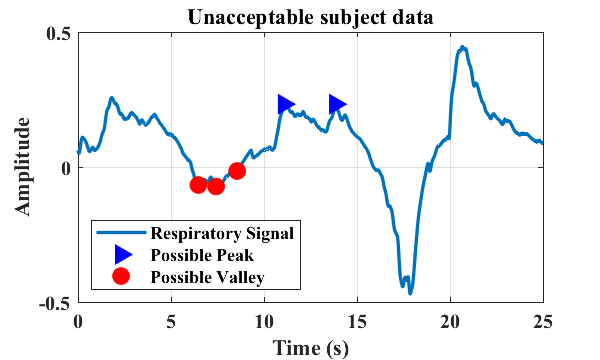}}
\caption{Example of unacceptable subject data in the CEBS database. Ground truth is impossible to establish, because there is uncertainty in the true locations of the peaks and valleys. Also, the atypical wave shape and unusually low frequency (0.1 Hz, versus the normal respiratory frequency range of 0.2-0.4 Hz) suggests that this measurement has been severely corrupted by motion artifact. Without ground truth labels, it is impossible to use this data to train or even evaluate our neural network.}
\label{bad_subject2}
\end{figure}

Likewise, for the CEBS database, some portions of the respiratory signals are too noisy for ground truth RR and \fit~values to be ascertained. As such, 25-second window snippets of noisy data similar to the one shown in Fig. \ref{bad_subject2} were excluded from the CEBS dataset.

\begin{figure}[htb]
\centerline{\includegraphics[scale=0.48]{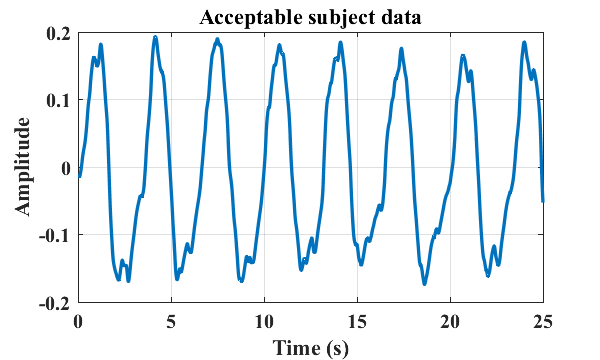}}
\caption{Example of acceptable subject data. The peaks and valleys are clearly defined, the wave shape resembles a typical respiratory signal, and the fundamental frequency falls within the normal range.}
\label{good_subject}
\end{figure}

\subsection{Annotation}
We developed a scheme to automatically annotate the reference respiratory phase classes from the respiratory signals. In this scheme, we first computed the moving average curve of the respiratory signal and used intercept pairs to locate the peaks \cite{Lu2006ASM}. We then performed edge detection \cite{FRIED20071063} to locate the valleys. These peak and valley locations were used to create the target class labels, $y$. The ground truth \fit~ is computed from these class labels using the duty cycle calculation explained in Section \ref{dcycle}.

\subsection{Training and Evaluation}
The MTL model was developed in Tensorflow Keras \cite{chollet2015keras}. The hyper-parameters used to train this network are shown in Table \ref{tab:NN_train}.
\begin{table}[htbp]
\caption{Neural Network Hyper-parameters}
\begin{center}
\begin{tabular}{|c|c|c|c|}
\hline
\textbf{Parameter} & \textbf{\textit{Value}}\\
\hline
Batch Size & 32\\
Training Optimizer & Adam \cite{Kingma2015AdamAM}\\
Learning rate & 0.001\\
Dropout & 0.2 \\
\hline
\end{tabular}
\label{tab:NN_train}
\end{center}
\end{table}
 The network is trained using one Tesla K80 GPU, with a 20\% cross-validation split of the training data. The model is then tested on a unique test set. This reserved test set is used to quantify the performance of both the neural network and the \fit~and respiratory rate extraction algorithm. The performance metrics reported for this work are the accuracy of the neural network classifier, the coefficient of determination ($\rm{R}^2$), and the root mean squared error (RMSE) of the estimated \fit~and RR. These are calculated as shown in Equations \ref{perform_metrics1} - \ref{perform_metrics2}  where TP, TN, FP and FN represent the number of true positive, true negative, false positive and false negative respectively. $\hat{\rm{p}}_{\rm{w}}$ is the estimated parameter value and $\rm{p}_w$ is the actual parameter value for each window w. The parameter value is either \fit~or RR
 \begin{eqnarray}
     {\rm{Accuracy}} = \frac{\rm{TP} + \rm{TN}}{\rm{TP}+\rm{TN}+\rm{FP}+\rm{FN}}
     \label{perform_metrics1}\\
     \rm{RMSE} = \sqrt{\frac{1}{n}\sum_{\rm{w}=1}^n\Big(\rm{p}_{\rm{w}} - \hat{\rm{p}}_{\rm{w}}\Big)^2}
     \label{perform_metrics2}
 \end{eqnarray}


\section{Results and Discussion}
\begin{figure}[ht]
\hspace*{-0.5cm}
\centerline{\includegraphics[scale=0.23]{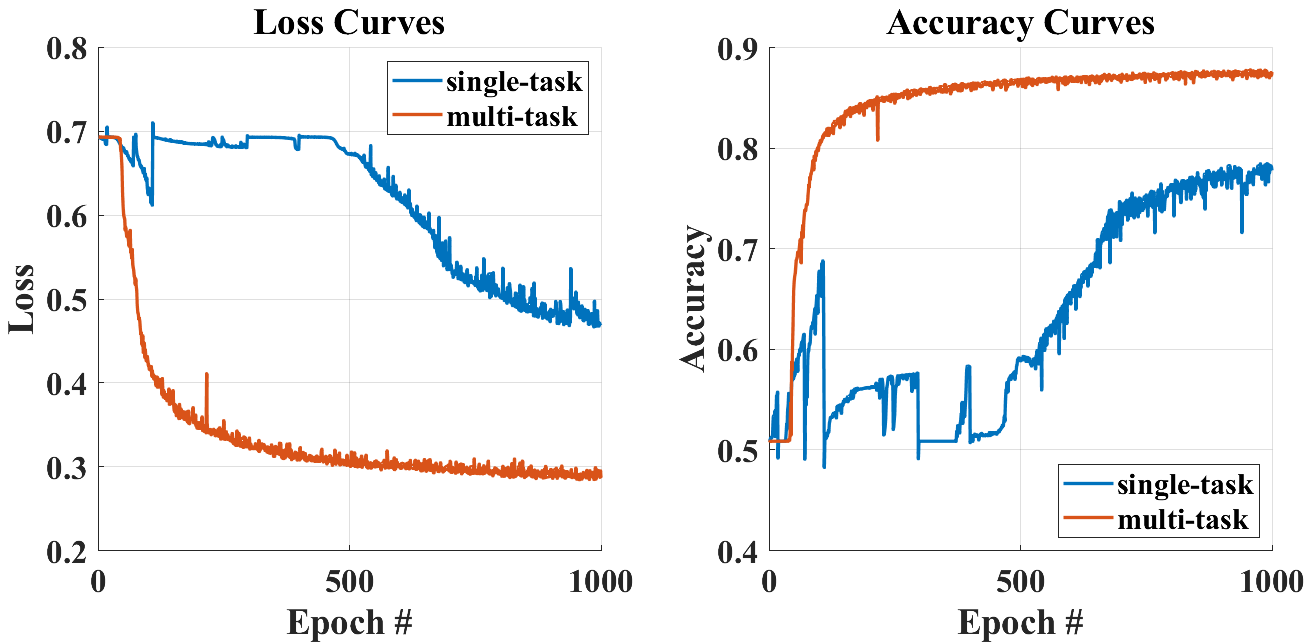}}
\caption{Plot of the loss and accuracy training profile of the neural network when evaluated on the validation set for the single-task and multi-task profiles.}
\label{jlearnvsnone}
\end{figure}

We studied the performance of the multi-task learning (MTL) algorithm described in Section \ref{MTL_sec} and compared it to a single-task binary classification algorithm (primary branch of the MTL). Fig. \ref{jlearnvsnone} compares the validation set's accuracy and loss curves over 1000 epochs.
A model checkpoint is implemented in both algorithms to save the weights corresponding to the model with the highest validation accuracy. The MTL algorithm reached the optimal training point 16 times faster than the single-task binary classification algorithm. Using the multi-task forces the model to learn features pertinent to the respiratory signal faster, thus reducing training time. The multi-task algorithm has 1656 parameters that utilize 96 kB memory, which, when pruned and quantized, can reduce the size by 35 times to about 3 kB without loss in accuracy \cite{Han2016DeepCC}, making its implementation on wearable devices plausible. 

The multi-task model outputs both the respiratory phase classifications ($y'$) and the respiratory signal ($z'$) from which the \fit~and RR are computed respectively.
The RR computed from the respiratory signal $z'$ is compared against the reference RR extracted from the true/ recorded respiratory signal $z$.

\subsection{MIMIC Results}
\begin{figure}[htb]
\centerline{\includegraphics[scale=0.52]{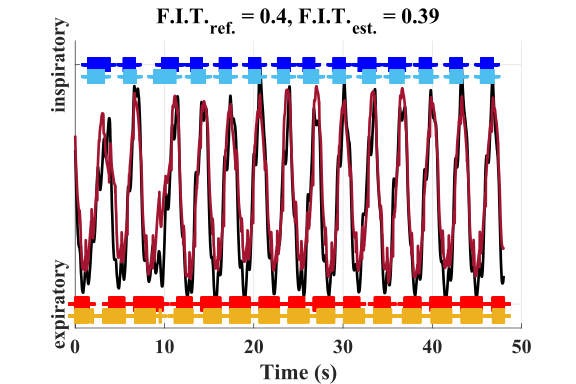}}
\caption{Plot of true (red and blue) vs predicted (gold and light blues) respiratory phases and true (black) vs predicted (burgundy) respiratory signal estimation of one window w1 in the test set}
\label{example_plot}
\end{figure}

The algorithm trained on the MIMIC training dataset is tested on the unique MIMIC test set. The $z'$ and $y'$ for each test window is extracted and compared to the true $z$ and $y$. Fig. \ref{example_plot} shows one test window output with the computed true and predicted \fit~values.

From this example, it is evident that the algorithm has the ability to extract the respiratory signal and respiratory phase classes from ECG MCL1 signals with a high degree of accuracy. The algorithm also estimates the \fit~with high precision as shown in the Fig. \ref{example_plot} whose estimated \fit~= 0.39, in comparison to the true \fit~= 0.4.
\begin{figure}[t!]
\centerline{\includegraphics[scale=0.512]{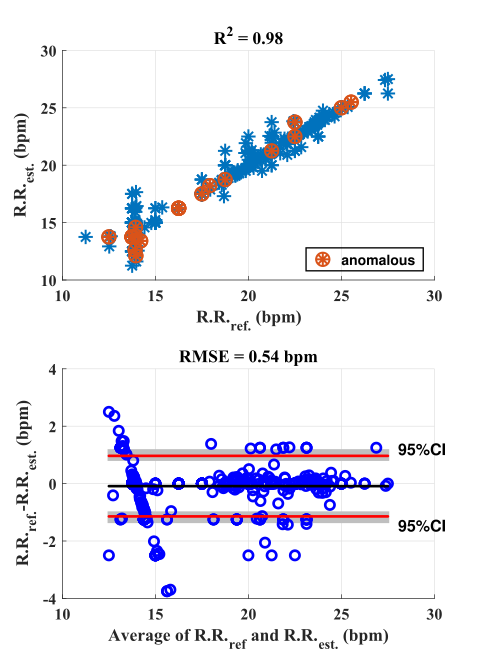}}
\caption{Scatter plot of true vs estimated respiratory rate for test subjects (top), Bland-Altman Plot of true and estimated respiratory rate for test subjects (bottom)}
\label{RR_plots}
\end{figure}
\begin{figure}[t!]
\centerline{\includegraphics[scale=0.512]{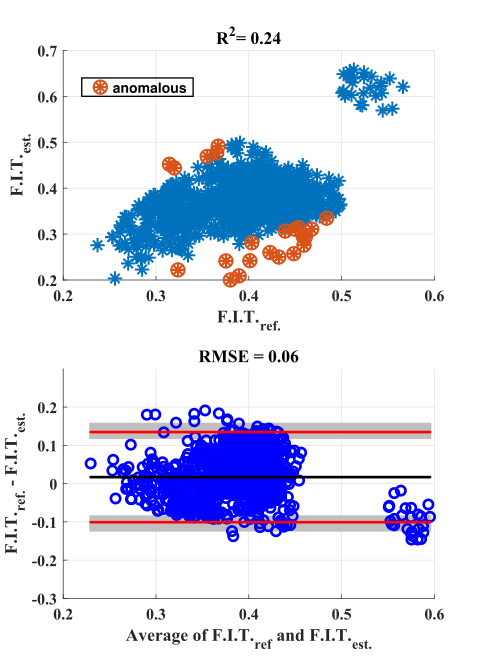}}
\caption{Scatter plot of true vs predicted \fit~for test subjects (top), Bland-Altman Plot of true and predicted \fit~for test subjects.}
\label{FIT_plots}
\end{figure}
The results on the entire MIMIC test set are compiled - the estimated respiratory rates are compared to the true/ reference respiratory rates, and the estimated \fit~are compared to the true/ reference \fit~. These results are displayed in the scatter plots and Bland-Altman plots shown in the Fig. \ref{RR_plots} for the respiratory rate and Fig. \ref{FIT_plots} for the \fit. 


Our algorithm extracts respiratory rate with an RMSE of 0.54 bpm and an $R^2$ value of 0.98 as shown by the RR scatter plot in Fig. \ref{RR_plots} - thereby outperforming the algorithm presented in \cite{Prinable2020DerivationOR} that reports a respiratory rate $R^2$ value of 0.87. Our respiratory rate performance is comparable to, and better than most existing techniques that have an $R^2 >$ 0.9. A comparison of our algorithm's effectiveness at extracting respiratory metrics against existing techniques in \cite{Prinable2020DerivationOR}, \cite{iPrext}, \cite{PCA}, \cite{fusion} and \cite{armband} is shown in Table \ref{tab:comparison}. 
\begin{table}[ht]
\caption{Comparison of the inspiratory and expiratory phase classification performance of this EDR algorithm and the EDR algorithm described in \cite{7408587} and comparison of the lung function metric estimation using this approach and approaches in \cite{Prinable2020DerivationOR, iPrext, PCA, armband}}
\begin{center}
\begin{tabular}{|c|c|c|c|c|c|c|c|}
\hline
\textbf{Work} & \multicolumn{2}{|c|}{\textbf{\fit/ I:E ratio}} & \multicolumn{2}{|c|}{\textbf{RR (bpm)}}\\
\cline{2-5} 
done &$\rm{R}^2$ & RMSE & $\rm{R}^2$ & RMSE \\
\hline
   \cite{Prinable2020DerivationOR} & -0.04 & - & 0.87 & - \\
\hline
    \cite{iPrext} & - & - & 0.99 & 0.03 \\
\hline      
    \cite{PCA} & - & - & 0.906 & 0.89 \\
\hline      
    \cite{armband} & - & - & 0.978 & -\\
\hline      
    \textbf{Our Work - MIMIC} & 0.24 &  0.06 & 0.98 & 0.54\\
\hline
    \textbf{Our Work - CEBS} & 0.62 &  0.11 & 0.9 & 0.66\\
\hline
\hline
\textbf{} & \textbf{Accuracy} \\
 &\textbf{(\%) of y} \\
\cline{1-2} 
   \cite{7408587} & 66.7\\
\cline{1-2} 
   \textbf{Our Work - MIMIC} & 70\\
\cline{1-2} 
   \textbf{Our Work - CEBS} & 75\\
\cline{1-2} 
\end{tabular}
\label{tab:comparison}
\end{center}
\end{table}

Our algorithm reports an RMSE of 0.06 and an $R^2$ value of 0.24 in the extraction of the \fit~as shown in the \fit~scatter plot in Fig. \ref{FIT_plots} - thereby also outperforming the algorithm presented in \cite{Prinable2020DerivationOR} that reports an $R^2$ value of -0.04. We compare the algorithm's performance at classifying the respiratory phases to another EDR algorithm developed in \cite{7408587} as shown in Table \ref{tab:comparison}. Our algorithm outperforms \cite{7408587}  by more than 4 percentage points in detecting the inspiratory and expiratory phases. Our algorithm is able to outperform the \fit~ and respiratory phase classification algorithms while maintaining exceptional performance in the computation of RR that is comparable to and even better than some existing techniques.
These results show a significant improvement in the ability to capture the subtle phase dependencies in the respiratory waveform from a single-lead ECG signal, in comparison to the algorithms presented by \cite{Prinable2020DerivationOR} and \cite{7408587}. This improvement shows that we can achieve even better accuracy in respiratory phase extraction and thus \fit~estimation, which will enable us to accurately infer lung airway obstruction.

\subsection{CEBS Results}
The CEBS database contains ECG Lead I signals instead of the ECG MCL1 signals in the MIMIC database. The MIMIC database is also over ten times larger than the CEBS database.
\begin{figure}[t!]
\centerline{\includegraphics[scale=0.512]{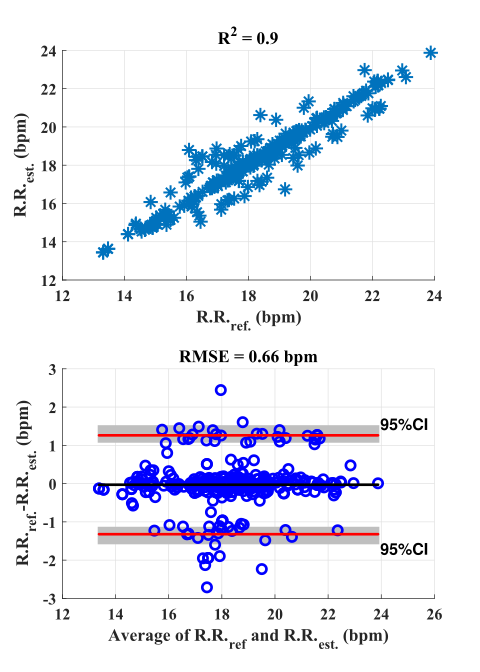}}
\caption{Scatter plot of true vs estimated respiratory rate for test subjects (top), Bland-Altman Plot of true and estimated respiratory rate for test subjects (bottom)}
\label{RR_plots_CEBS}
\end{figure}
\begin{figure}[t!]
\centerline{\includegraphics[scale=0.512]{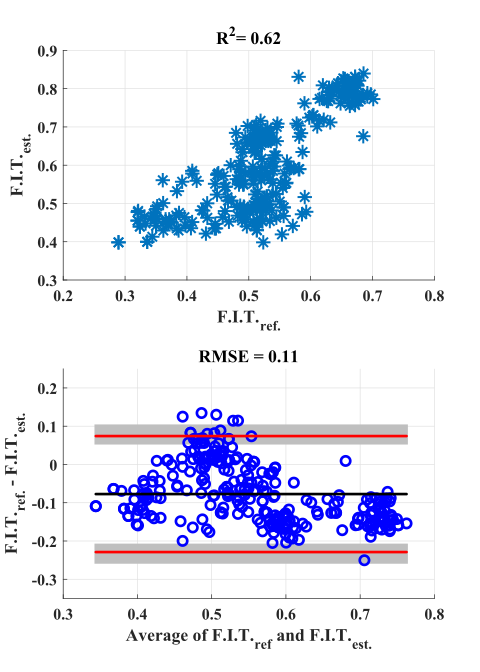}}
\caption{Scatter plot of true vs predicted \fit~for test subjects (top), Bland-Altman Plot of true and predicted \fit~for test subjects.}
\label{FIT_plots_CEBS}
\end{figure}
We take advantage of this difference in databases and apply transfer learning from the pre-trained MIMIC algorithm to fine-tune on the CEBS training dataset. The learned, fine-tuned algorithm is then evaluated on the unique CEBS test dataset. 
The scatter plot and the Bland-Altman plot results for the RR and \fit~estimation on the CEBS test dataset are shown in Figs. \ref{RR_plots_CEBS} and \ref{FIT_plots_CEBS}. Our algorithm maintains exceptional performance in extracting the RR with an $R^2$ value of 0.9 and an RMSE of 0.66 bpm. The \fit~performance on the CEBS dataset is even better than that on the MIMIC dataset with an $R^2$ value of 0.62 and an $RMSE$ of 0.11. We believe that this performance increase is due to the guaranteed temporal alignment of the CEBS dataset unlike the MIMIC dataset \cite{mimic_corr}. It is our belief that the unknown and inconsistent delays/ shifts between the respiratory and ECG waveforms (even within a subject) in the MIMIC dataset, make the learning task very difficulty, and even more so, make the evaluation of the algorithm challenging. The  MIMIC dataset, however, is a large and can be used to pre-train a network especially in cases with limited temporally-aligned data.

\section{Limitations}
\begin{figure}[htb]
\centerline{\includegraphics[scale=0.512]{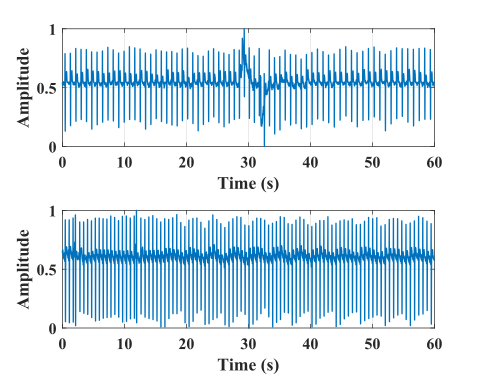}}
\caption{Anomalous ECG signals that report poor \fit~estimation performance with transient breaks that alter the dynamic range of the ECG signals (top) and irregular ECG time interval patterns (top and bottom).}
\label{anomalous_ecg}
\end{figure}

This algorithm has one main limitation as shown by the anomalous points in Figs. \ref{RR_plots} and \ref{FIT_plots}; the algorithm performs poorly on irregular / anomalous ECG data. Of the 982 windows, 22 had an \fit~estimation absolute error that was greater than 30\%. These 22 subjects either had irregular ECG signals or transient breaks that altered the dynamic range of the ECG signals as shown in Fig. \ref{anomalous_ecg}. Because we are using the ECG signals as direct inputs to the algorithm, differences in the ECG signals like changes in the dynamic range of the ECG signal affect the output of the algorithm. In the future we want to explore using generative adversarial networks to augment our ECG signal data augmentation. The data augmentation will include distortions to normal ECG signals that will change the dynamic range and regularity in time interval patterns of normal ECG signals while maintaining the amplitude modulation, and QRS complex relationships. The goal is to try and make the model more robust to anomalous ECG signals.

\section{Benefits of this approach}
\begin{figure*}[htb]
\centerline{\includegraphics[scale=0.55]{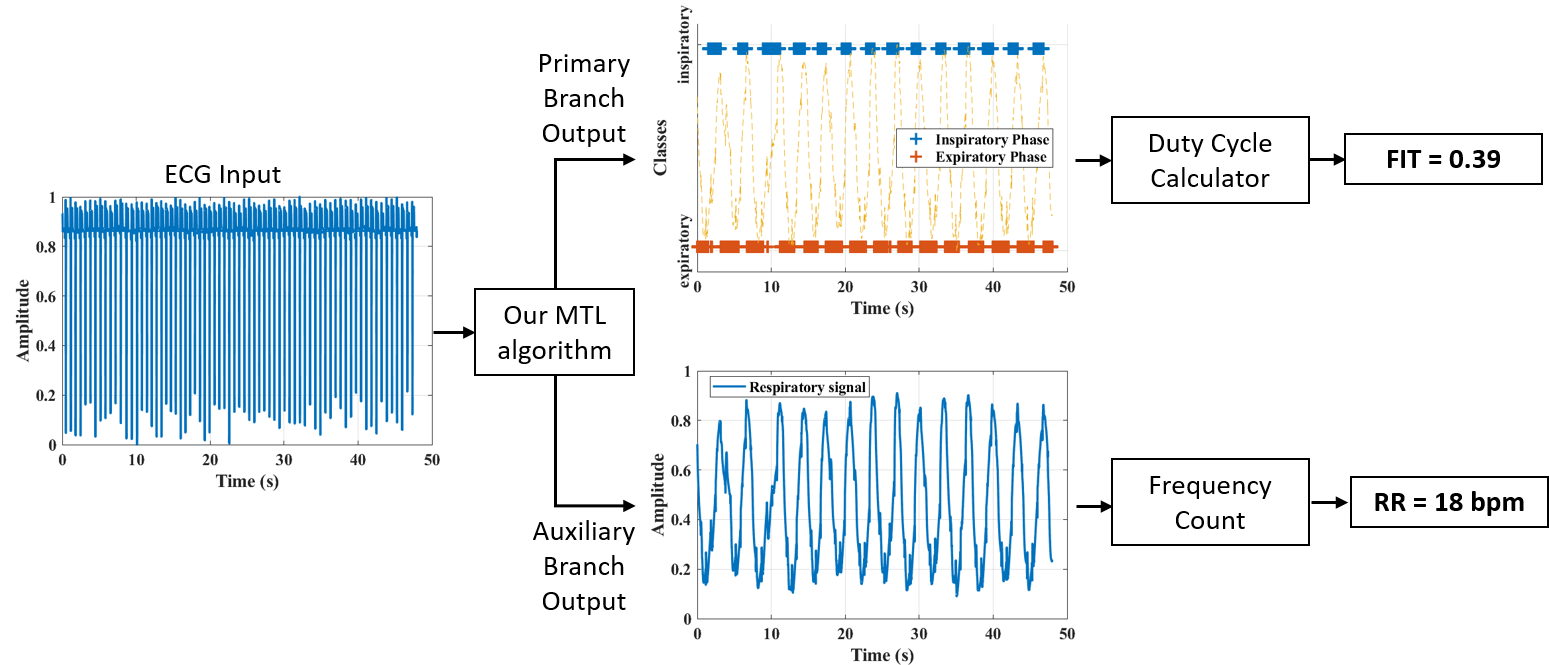}}
\caption{Overall system design showing the ECG signal input to the MTL model, and its outputs from the primary and auxiliary branches with the computed \fit~ and RR}
\label{system_design}
\end{figure*}
\begin{figure}[htbp]
\centerline{\includegraphics[scale=0.5]{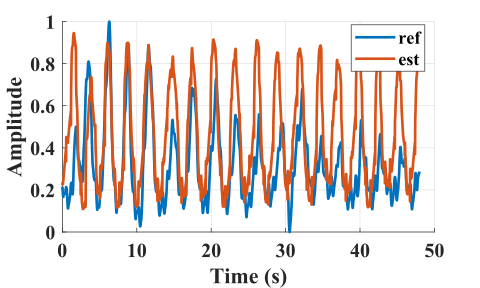}}
\caption{Respiratory signal extracted from single lead ECG signal compared to its reference respiratory signal with diminishing amplitude.}
\label{shallow_resp}
\end{figure}

The algorithm is able to accurately estimate most respiratory signals, even those reference respiratory signals whose amplitude is reducing, as shown in Fig. \ref{shallow_resp}.

Additionally, this MTL model utilizes a single lead ECG (ECG MCL1 or ECG Lead I) which is measured by ubiquitous wearable devices, favoring its applicability to current wearable devices. This MTL model not only extracts the respiratory signal, but also computes the respiratory rate (RR) and fractional inspiratory time (\fit). It performs these computations on streams of data negating the need for vast amounts of memory, and providing the opportunity to observe trends in RR and \fit~over time. This ability can open the door to more informed studies on changes in RR and \fit~over time. This approach also reports better accuracy than the limited existing approaches mentioned (\cite{Prinable2020DerivationOR, 7408587}).

\section{Conclusion}
Our paper proposes a novel approach to extract \fit~and respiratory rate from a single ECG signal as shown in Fig. \ref{system_design}. Our algorithm is demonstrated to have significantly higher respiratory phase classification accuracy, and \fit~estimation, in comparison to existing techniques, and exceptional respiratory rate estimation. It can also be implemented on a wearable device. Our algorithm would increase ECG signals' utility and motivate continuous monitoring of pulmonary parameters from prevalent wrist-worn wearable devices with built-in ECG sensors.

\bibliographystyle{./IEEEtran}
\bibliography{./FIT_ECG}



\end{document}